\documentclass[sigconf]{acmart} 
\setcopyright{none}
\renewcommand\footnotetextcopyrightpermission[1]{}
\usepackage{pifont}
\newcommand{\cmark}{\ding{51}}
\newcommand{\xmark}{\ding{56}}

  \usepackage{booktabs}
  \usepackage{multirow}
  \usepackage{array}
  \usepackage{xcolor}
  \usepackage{svg}
  \usepackage{colortbl}
  \usepackage[table]{xcolor}
  \usepackage{makecell}
\AtBeginDocument{%
  }

\setcopyright{none}
\renewcommand\footnotetextcopyrightpermission[1]{}
\begin{document}

\title{GlassTENG: Self-Powered Triboelectric Nanogenerator based Sensing of Pulse, Jaw, and Upper Facial Activity 
from Everyday~Glasses}


\author{Raj N. Dave}
\authornote{These authors contributed equally.}
\email{rajdave2027@u.northwestern.edu}
\affiliation{
  \institution{VAK Embodied Systems Lab, Northwestern University}
  \city{Evanston}
  \state{Illinois}
  \country{USA}
}

\author{Jovanis Prodanich}
\authornotemark[1]
\email{jovanisprodanich2026@u.northwestern.edu}
\affiliation{
  \institution{VAK Embodied Systems Lab, Northwestern University}
  \city{Evanston}
  \state{Illinois}
  \country{USA}
}

\author{Yung-ching Lai}
\authornotemark[1]
\email{yungching.lai@u.northwestern.edu}
\affiliation{
  \institution{VAK Embodied Systems Lab, Northwestern University}
  \city{Evanston}
  \state{Illinois}
  \country{USA}
}

\author{Oscar Jakacki}
\authornote{These authors contributed equally.}
\email{oscarjakacki2028@u.northwestern.edu}
\affiliation{
  \institution{VAK Embodied Systems Lab, Northwestern University}
  \city{Evanston}
  \state{Illinois}
  \country{USA}
}

\author{Stanley Lin}
\authornotemark[2]
\email{stanleylin2028@u.northwestern.edu}
\affiliation{
  \institution{VAK Embodied Systems Lab, Northwestern University}
  \city{Evanston}
  \state{Illinois}
  \country{USA}
}

\author{Jack Thoene}
\email{jackthoene2025@u.northwestern.edu}
\affiliation{
  \institution{VAK Embodied Systems Lab, Northwestern University}
  \city{Evanston}
  \state{Illinois}
  \country{USA}
}

\author{Nabil Alshurafa}
\email{nabil@northwestern.edu}
\affiliation{
  \institution{Northwestern University}
  \city{Evanston}
  \state{Illinois}
  \country{USA}
}

\author{Nivedita Arora}
\email{nivedita@northwestern.edu}
\affiliation{
  \institution{VAK Embodied Systems Lab, Northwestern University}
  \city{Evanston}
  \state{Illinois}
  \country{USA}
}


\begin{abstract}

Smart glasses maintain near-continuous skin contact at multiple arterial and muscular sites, making them a promising platform for physiological sensing. In practice, though, two factors make sustained daily wear and longitudinal deployment impractical for the quantified self: the discomfort of prolonged sensor–skin contact (e.g., gels and adhesives) and the sensor power demands that increase battery size, weight, and maintenance burden. We present GlassTENG, an ultra-low-power sensor that embeds three custom-fabricated triboelectric nanogenerators (TENGs) into a glasses frame
at the angular artery on the nasal bridge, the superficial temporal artery on an extended arm, and the temporalis muscle at the temple. Each GlassTENG sensor is self-powered in transducing mechanical energy to electrical energy and consumes 1.36\,$\mu$W per sensor at the analog front-end. GlassTENG enables simultaneous capture of arterial pulse waveforms, jaw kinematics (e.g., clenching, tapping, eating), and upper facial activity (e.g., blinking, eyebrow movement). In a 20-participant user study, we achieve 93.8\% accuracy across six jaw and upper facial activities and estimate heart rate with a mean absolute error of 1.82 beats per minute (BPM) relative to a ground-truth chest-strap sensor in 30s windows. Together, these results establish a future pathway toward a longitudinally worn, ultra-low-power, glasses-based physiological monitoring platform.
\end{abstract}


\begin{teaserfigure}
\vspace{-0.2in}
  \includegraphics[width=\textwidth]{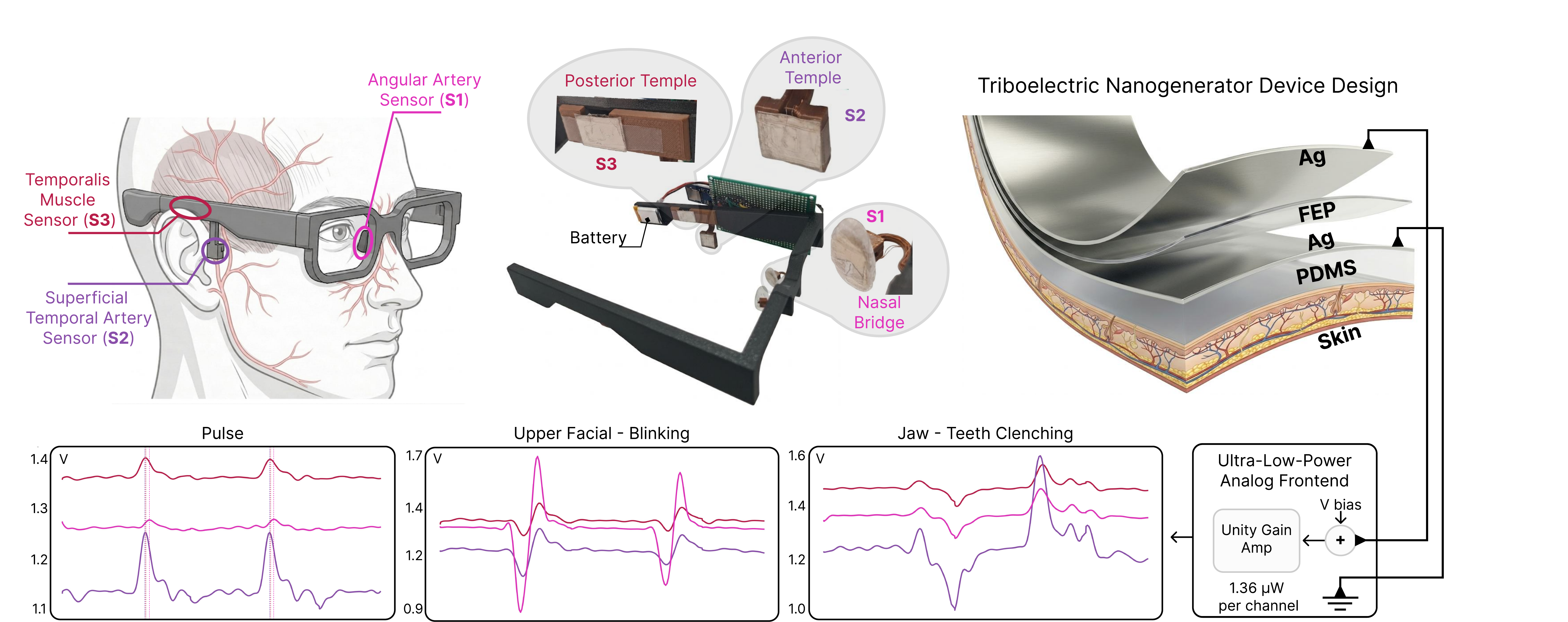}
  \caption{GlassTENG embeds three custom-fabricated Triboelectric Nanogenerator (TENG) sensors into everyday glasses at three facial sites that are physiologically rich: nasal bridge, posterior temple, anterior temple . A 1.36$\mu$W-per-channel high impedance analog front-end conditions the TENG's voltage output, enabling simultaneous recovery of arterial pulse waveforms and six upper facial and jaw activity.} 
  \Description{temp fig 1}
  \label{fig:temp_teaser}
\end{teaserfigure}

\maketitle

\pagestyle{plain}

\section{Introduction} \label{sec:intro}

The human face is a remarkably rich environment, offering varied
anatomical vantage points to capture vital signs and complex
actions such as heart rate~\cite{glabella}, jaw
kinematics~\cite{facial_emg_nature,facial_emg_ex2}, eye
movements~\cite{eyegesturelistener}, and facial
expressions~\cite{emotionglasses}. 
Everyday eyeglasses provide a convenient interface to capture these signals: worn daily by over 4 billion people~\cite{visioncouncil2021,who2019vision} with demand still rising~\cite{Subin2026}, they are a habitual facial accessory that requires no behavioral change. While this makes glasses preferable to fitness bands for long-term adoption~\cite{wearable_abandonment}, realizing the potential of glasses as a longitudinal bio-sensing platform imposes a hard constraint: \textit{they must remain glasses in form factor, weight, and user experience.}

Electronic weight, bulk, and maintenance erode the unobtrusiveness users have already accepted into their lives, and a single component largely drives all three: the battery. In wearables, the battery dominates weight,~\cite{battery_bottle_neck}, dictates bulk, imposes the burden of recharging, and has been a leading cause of abandonment~\cite{batt_recharge, starner2002challenges}. This raises an important research question: can smart glasses support continuous rich physiological sensing while remaining lightweight, unobtrusive, and effectively maintenance-free? For users who already wear glasses, physiological sensing should become an invisible capability requiring no additional maintenance or behavioral change. Realizing this vision means eliminating the battery in favor of energy harvesters and, in turn, \textbf{rethinking the sensing architecture of smart eyewear so that it can operate within an extremely constrained energy-harvested power budget}~\cite{internet_of_batteryless_things}.


To understand how constrained that 
budget truly is, we calculate an estimate of how much a glass frame can realistically harvest. Given that people wear glasses primarily in lit environments, an everyday glass frame covered with solar cells in indoor light (40--60\,cm\textsuperscript{2} exterior area, 5$\mu$W/cm\textsuperscript{2}) harvests 200--300$\mu$W. Other energy harvesters, like body heat and electromagnetic movement, produce much less power than what photodiodes provide \cite{indoor_solar_harvesting, energy_harvesing_grosse}. The power generated can barely power a modern duty-cycled low-power micro-controller (ASIC~\cite{vanhelleputte2015soc}) set for continuous data acquisition, even after decades of research in battery-free/low-power IoT~\cite{internet_of_batteryless_things}. This leaves only few microwatts of power for physiological-to-electrical transduction and its analog front-end. \textbf{This paper asks whether rich physiological sensing, including arterial pulse, oculofacial muscle activity, and jaw motion, can be achieved within a sensing power budget of only 5$\mu$W on a glasses platform.} 

Current sensing modalities used in smart eyewear, including optical Photoplethysmography (PPG),  Electromyography (EMG)~\cite{biteglasses,dieteyeglasses}, load cells~\cite{glassense}, and
photo-reflective arrays~\cite{photoreflective}, each consume milliwatts
continuously, exceeding this budget by more than an order of magnitude
(Table~\ref{tab:sensor_comparison}). The need is of ``self-powered sensors'': transducers that harvest the energy to sense directly from the phenomenon being sensed. This is a natural fit for facial activity, since many physiological signals of interest e.g., pulse and jaw muscle movement are themselves small mechanical events. The central challenge is whether a self-powered sensor can be made sensitive enough to convert these faint movements into usable electrical signals without drawing any power. 

We present, GlassTENG, a set of custom-fabricated flexible self-powered triboelectric nanogenerator (TENG) sensors, optimized for the tiny biomechanical force regime of arterial pulse and facial muscle activity. The sensors are integrated
into a standard glasses frame at three anatomically targeted sites --- the
nasal bridge, anterior temple, and the posterior temple (Figure~\ref{fig:temp_teaser}), where they convert tiny-mechanical forces from arterial pulse and facial muscle movement directly into electrical signals. GlassTENG is supported by an ultra-low-power analog front-end consuming less than $< 5uW$ while maintaining comfortable skin contact. This paper focuses on the sensing front-end and physiological signal acquisition; energy harvesting and fully battery-free end-to-end integration remain future work. Our contributions are as follows:
\setlength{\leftmargini}{1.1em}\begin{itemize}
\item \textbf{Sensor design, fabrication and glass integration.} Custom multi-layer (PDMS/Ag and FEP/Ag) TENG sensors, sensitive to tiny forces (0.01N -- 2N) generated by facial physiological activity e.g., pulse, jaw (50mV -- 0.8V), optimized for skin-conforming contact at three anatomically distinct sites (angular artery, superficial temporal artery, and temporalis muscle) on glasses. 

\item \textbf{Ultra-low-power analog front-end} A custom high-impedance (18T$\Omega$) analog front-end matched to each GlassTENG sensor's unique mechanically induced voltage output, achieving a power budget of 1.36$\mu$W per channel (4.1\,$\mu$W total).

  

\item \textbf{Multi-modal physiological sensing.}  Consistent arterial pulse from all three sites, achieving mean absoloute error (MAE) of \textbf{1.82} for heart rate against ground truth (opening a pathway to cuffless blood pressure) and 93.8\% accuracy across \textbf{6} facial activity classes, validated through a 20-participant user study.

 \end{itemize} 
\section{Related Work} \label{sec: rel_works}
Realizing glasses-based physiological sensors suitable for longitudinal
deployment requires balancing three design goals: ultra-low power consumption
(a few microwatts), comfortable sensors (skin contact free of gels and
adhesives), and rich sensing capability spanning multiple physiological
signals. We organize the literature review around these goals, covering
physiological eyewear sensing approaches and methods to make sensors
self-powered.



\definecolor{lightred}{RGB}{255,180,180}
\definecolor{lightgreen}{RGB}{180,230,180}

\begin{table}[t]
\centering
\caption{Power constraints for physiological eyewear sensing
}
\vspace{-3mm}
\label{tab:sensor_comparison}
\scriptsize
\setlength{\tabcolsep}{3.0pt}
\renewcommand{\arraystretch}{1.25}
\begin{tabular}{l r c c c}
\toprule
\textbf{Sensor}
  & \textbf{\makecell{front-end\\ Power}}
  & \textbf{\makecell{Sustained\\ Wearability}}
  & \textbf{Signals Detected} \\
\midrule
Opt. PPG
  & \cellcolor{lightred}1--10\,mW\textsuperscript{*}
  & \cmark 
  & Pulse, HR~\cite{glabella,pulseglasses}; emotion~\cite{emotionglasses} \\
EDA
  & \cellcolor{lightred}3--5mW\textsuperscript{*}
  & \cmark 
  & Emotion~\cite{emotionglasses} \\
Surface EMG
  & \cellcolor{lightred}10--160\,mW\textsuperscript{*}
  & \makecell{\xmark  (gel required)}
  & Chewing~\cite{dieteyeglasses,biteglasses,freelivingchewing} \\
Load Cell
  & \cellcolor{lightred}1--30mW\textsuperscript{*}
  & \cmark 
  & Jaw, food intake~\cite{glassense,foodintakesmartglasses} \\
Piezo + IMU
  & \cellcolor{lightred}3--12mW\textsuperscript{*}
  & \makecell{\xmark \scriptsize adhesive required} 
  & Jaw, expressions~\cite{meciface} \\
Opt. Tracking
  & \cellcolor{lightred}5--15mW\textsuperscript{*}
  & \cmark 
  & Chewing, clenching~\cite{opticalchewing} \\
Microphone
  & \cellcolor{lightred}$\sim$2--42\,mW\textsuperscript{*}
  & \cmark 
  & Chewing~\cite{chewingrazor}; activity~\cite{actsonic} \\
Photo-reflect.
  & \cellcolor{lightred}6--10mW\textsuperscript{*}
  & \cmark
  & Expressions~\cite{photoreflective} \\
\midrule
\textbf{GlassTENG (ours)}
  & \cellcolor{lightgreen}\textbf{4.1\,$\mu$W}
  & \cmark 
  & \textbf{Pulse (3 sites), jaw, upper facial} \\
\bottomrule
\multicolumn{5}{p{0.88\columnwidth}}{\scriptsize
Sustained Wearability  (no maintenance, gel-free, long term daily usage);
\textsuperscript{*}~Estimated front-end power values based on components used
}\vspace{-5mm}
\end{tabular}
\end{table}

\begin{figure*}[t]
    \centering
    \includegraphics[width=2.1\columnwidth]{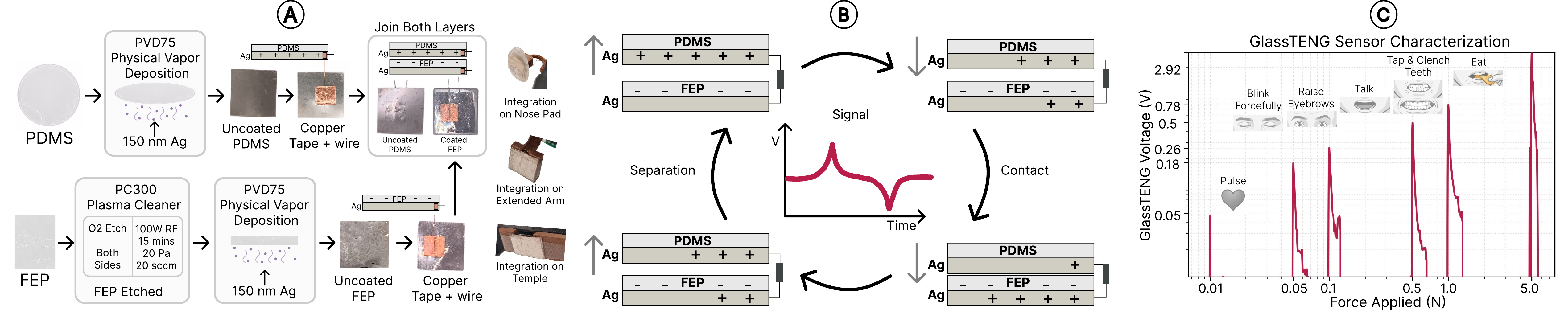}
    \caption{(A) Fabrication process of the triboelectric nanogenerator (TENG) sensor. (B) Working principle of the vertical contact separation TENG. (C) Sensor characterization plot with reference activities and respective voltages.}
    \label{fig:fabrication_working_characterization}
    \vspace{-3mm}
\end{figure*}

\vspace{-0.1in}
\subsection{Physiological Sensing from Eyewear} \label{sec:2.1}
\textbf{Cardiovascular Sensing:}
 Continuous pulse monitoring via wrist PPG is widely deployed and used in smart watches. The face offers a compelling alternative for cardiovascular sensing. The foundational work by Hertzman et. al in  \cite{hertzman1937} established that arterial pulse is optically accessible from the nasal area. Subsequent glasses-based systems demonstrated  multiple applications like blood pressure proxies and heart rate monitoring ~\cite{clip_free_glabella_predecesor,glabella,pulseglasses}. PPG has also been fused with Electrodermal Activity (EDA) for emotion sensing~\cite{emotionglasses}.  However, all these systems rely on continuously powered optical emitters, which strain both battery life and form factor ~\cite{application_hypertension_using_ppg,BP_based_on_PTT_PAT_review,ppg_limitations}. GlassTENG addresses these limitations by lowering the power requirement for cardiovascular sensing.

\noindent\textbf{Jaw and Upper Facial Activity Sensing:} Eyeglasses have been used to sense both upper-face and jaw activity. For jaw activity, prior work has used EMG on the temporalis muscle for food-type classification and chewing detection~\cite{dieteyeglasses,biteglasses,freelivingchewing}, load cells for jaw rotation~\cite{glassense,foodintakesmartglasses}, and mechanomyography, acoustic, and optical methods for eating and expression recognition~\cite{meciface,chewingrazor,opticalchewing}. For upper-face activity, blink rate and eyebrow raises have been captured via IR reflection, acoustic vibration, and laser interferometry~\cite{photoreflective,acousticfacial,non_table_tunnelvision,eyegesturelistener,non_table_capglasses}. Blink patterns and eyebrow raises have been leveraged for fatigue monitoring, cognitive load estimation, and hands-free interaction~\cite{blink_and_cognitive_load,eye_and_fatigue_nature,eyebrow_as_communication}. Across these approaches, electrode maintenance and power-hungry front-end hardware compromise wearability. Table~\ref{tab:sensor_comparison} quantifies these limitations across modalities. GlassTENG captures both jaw and upper-face signals with an ultra-low-power front-end design, creating a pathway toward the development of longitudinally deployed platforms for physiological sensing.

\vspace{-3mm}
\subsection{Self-Powered and Battery Free Sensing} \label{sec:2.2}
Battery life remains one of the primary barriers to long-term wearable adoption~\cite{wearable_abandonment}. A growing body of work has explored self-powered and battery-free sensing with extremely low-power micro-controllers and custom silicon chips \cite{vanhelleputte2015soc, msp430-datasheet}. Energy harvesting approaches, including triboelectric~\cite{self_powered_PDms}, piezoelectric~\cite{self_power_piezo}, and thermoelectric~\cite{self_power_thermo} mechanisms, have been used to convert ambient mechanical or thermal energy to usable signals. Prior works have demonstrated battery-free glasses for applications like eye tracking and video streaming \cite{battery_free_eye_glasses,battery_free_hd_streaming}. GlassTENG builds on this principle of self-powered wearables and lays the foundation for future battery-free operation for sensing facial and cardiovascular activity.


\vspace{-3mm}
\subsection{Triboelectric Nanogenerator-Based Sensing} \label{sec:2.3}

Triboelectric nanogenerators have been demonstrated to have self-powering capability \cite{teng_original, Teng_new_energy, saturn_original, mars_original}. Furthermore, TENGs can be made with common materials from daily life such as paper, fabrics, PTFE, PDMS, etc. making them ideal for simple fabrication of self powered sensors \cite{teng_sp_new_energy, teng_progress, texteng}. They have been applied various applications like wrist-pulse sensing, respiration, muscle activity, and gait analysis in wearable contexts~\cite{teng_original,teng_wrist_pulse,teng_respiration,teng_muscle,gait_teng}. High-sensitivity TENG materials have demonstrated a wide range of applications from machine monitoring \cite{teng_for_machine_monitoring} to sufficient resolution for arterial pulsation ~\cite{teng_nanopore_pulse}. Recently, Qu et. al ~\cite{qu2025} integrated a TENG into a glasses frame primarily for hemifacial spasm detection, incidentally capturing pulse as a secondary result. GlassTENG exploits all the advantages provided by TENGs like high sensitivity, easy fabrication, and self-powering capability while providing an ultra-low-power sensing front-end with potential use in sustained wearability applications.

\section{GlassTENG System Design} \label{sec:glassteng_system_design}

This section describes the co-design of sensor materials, device design, analog front-end, and glasses integration — iteratively refined together to capture physiological signals from the face. 
\vspace{-3mm}
\subsection{Sensor Design,  Fabrication and Working} \label{sec:design_fabrication}
\noindent \textbf{Sensor Design (Figure \ref{fig:temp_teaser}).} Each GlassTENG sensor consists of a two-layer triboelectric nanogenerator architecture comprising polydimethylsiloxane (PDMS) and fluorinated ethylene propylene (FEP), with thin silver films serving as electrodes on each layer. 

\noindent \textbf{Fabrication (Figure \ref{fig:fabrication_working_characterization}A).} \textbf{i)} \textit{PDMS Sheet:} PDMS base and curing agent (Sylgard 184, Dow Inc) were mixed at a 10:1 weight ratio, and degassed under a vacuum to remove entrained air bubbles. The mixture was cast into a flat mold, spin-coated to ensure uniform thickness ($\sim$490$\mu$m), and cured overnight at room temperature to yield flexible elastomeric films.  \textbf{ii)} \textit{FEP Sheet and Etching:} Both sides of the FEP film were subjected to oxygen plasma etching prior to electrode deposition to enhance the triboelectric surface charge density. Initial 
prototypes etched only one FEP surface; extending the process to both sides measurably increased signal amplitude, motivating the bilateral etching. Etching parameters were: RF power of 100~W, O\textsubscript{2} flow rate of 
20~sccm, chamber pressure of 20~Pa, and etch duration of 15~minutes per side.  This produced a roughened morphology that substantially increased the effective 
contact area and charge generation upon repeated contact with PDMS. \textbf{iii)} \textit{Physical vapor 
deposition (PVD):} PVD was then used to deposit 150~nm conformal silver electrodes onto 
both films. \textbf{iv)} \textit{Electrical connections} were established by affixing copper tape to the exposed silver electrode surfaces along with a 38~AWG enameled copper wire. \textbf{v)} \textit{Assembly:} Both layers were stacked and laminated with adhesive tape such that bare PDMS faces the skin, and the Ag-coated PDMS and etched FEP surfaces face each other to form the triboelectric interface. This prevents interference from skin oils. The assembled sensors are then integrated into the glasses frame. 

\noindent \textbf{Working Principle:} The GlassTENG sensors operates on vertical 
contact-separation, as illustrated in 
Figure~\ref{fig:fabrication_working_characterization}B. When compression from 
muscle movement or arterial pulse deforms the PDMS layer, the Ag electrode on 
PDMS contacts the FEP film. Due to FEP's higher electron affinity, electrons 
transfer from Ag to FEP, leaving FEP negatively charged and the PDMS-side Ag 
positively charged. Subsequent relaxation separates the layers, inducing a 
potential difference across the electrodes and driving current through the external 
circuit. This produces 
an alternating signal whose amplitude reflects the magnitude of the applied force.

\vspace{-3mm}
\subsection{Sensor Placement and Glass Integration} \label{sec:sensor_placement}
GlassTENG integrates three sensors (S1, S2, S3) into the glasses frame (Figure~\ref{fig:temp_teaser}), each positioned over a distinct anatomical site
selected to capture specific biomechanical signals.
 
\noindent \textbf{Angular Artery Sensor (S1 -- Nasal Bridge ):}
The nasal bridge offers access to a physiologically rich site, making pulse and eye movements accessible through the angular artery and the oculofacial muscles \cite{glabella,selfpoweredeye}. S1 is positioned on the angular artery (through the nasal bridge), making it sensitive to both vascular pressure and deformation from surrounding oculofacial muscles. Early prototypes used 3D-printed flat nose pads, which produced inconsistent skin contact while testing; replacing these with adjustable off-the-shelf nose pads from commercial eyewear yielded a comfortable fit across varying nasal structures. Beyond pulse, S1 captures upper facial activity including forceful blinks and eyebrow movements.
 
\noindent \textbf{Superficial Temporal Artery Sensor (S2 -- Anterior Temple):}
S2 targets the superficial temporal artery, a highly accessible site for
non-invasive pulse sensing on the face \cite{glabella,qu2025}. It is mounted on a dedicated arm
extending downward from the frame, maintaining consistent pressure against the skin
without compromising comfort. S2 serves as a source for masticatory
signals, capturing teeth clenching, teeth tapping, and the rhythmic mechanics of chewing.

\noindent \textbf{Temporalis Muscle Sensor (S3 -- Posterior Temple ):}
S3 is situated at the posterior temple, directly over the temporalis muscle and the distal extension of the superficial temporal artery, making it a rich point for masticatory movements and pulse detection \cite{glabella, dieteyeglasses, freelivingchewing}. It is embedded into the temple arm just above the ear, as this position benefits from the natural clamping force of the glasses frame to ensure consistent contact and signal stability. S3 provides a complementary jaw signal to S2, together enabling characterization of complex oral actions.

\noindent With placement sites established, we characterized each sensor's force-voltage 
response to verify that the physiological forces at these sites produce 
measurable, distinguishable output voltages.

\vspace{-3mm}

\subsection{Sensor Characterization}\label{sec:characterization}

To characterize GlassTENG's sensitivity, we applied known forces to the sensor using a force gauge and recorded the resulting peak open-circuit voltages via a high-impedance (1~G$\Omega$) NI-DAQ . Rather than compressing the sensor directly, forces were removed from the sensor, triggering contact-separation and generating a voltage pulse that mirrors the 
transient nature of physiological signals in actual use. Simultaneously, a NI-DAQ was used to record from GlassTENG sensors integrated into the glasses  (Section~\ref{sec:sensor_placement}) as participants performed different activities. Comparing both force guage and GlassTENG datasets allowed us to map force, voltage, and biomechanical signals to one another (Figure~\ref{fig:fabrication_working_characterization}C). 

\noindent The sensor exhibits a nonlinear but monotonically increasing voltage response across 0.01--5~N, providing sufficient dynamic range to span all target activities. Arterial pulse (0.01~N) produces 50--100~mV; oculofacial activities such as blinks and eyebrow raises (0.05--0.1~N) yield 200--500~mV; and jaw 
activities including clenching and chewing (0.5--2~N) reach up to 1.5~V. These values vary across participants due to differences in facial anatomy and muscle strength, motivating per-participant normalization during classification. This characterization directly informed the analog front-end design: S3, routinely encountering the higher force regime, uses a voltage divider to prevent clipping within the 3.3~V rail, while S1 and S2 use the full unity-gain path.

\subsection{Analog Front-End Circuit Design} \label{sec:analog_front-end}

\vspace{-4mm}
\begin{figure}[!ht]
  \centering
  \includegraphics[width=0.8\columnwidth, trim=0 0 0 100pt, clip]{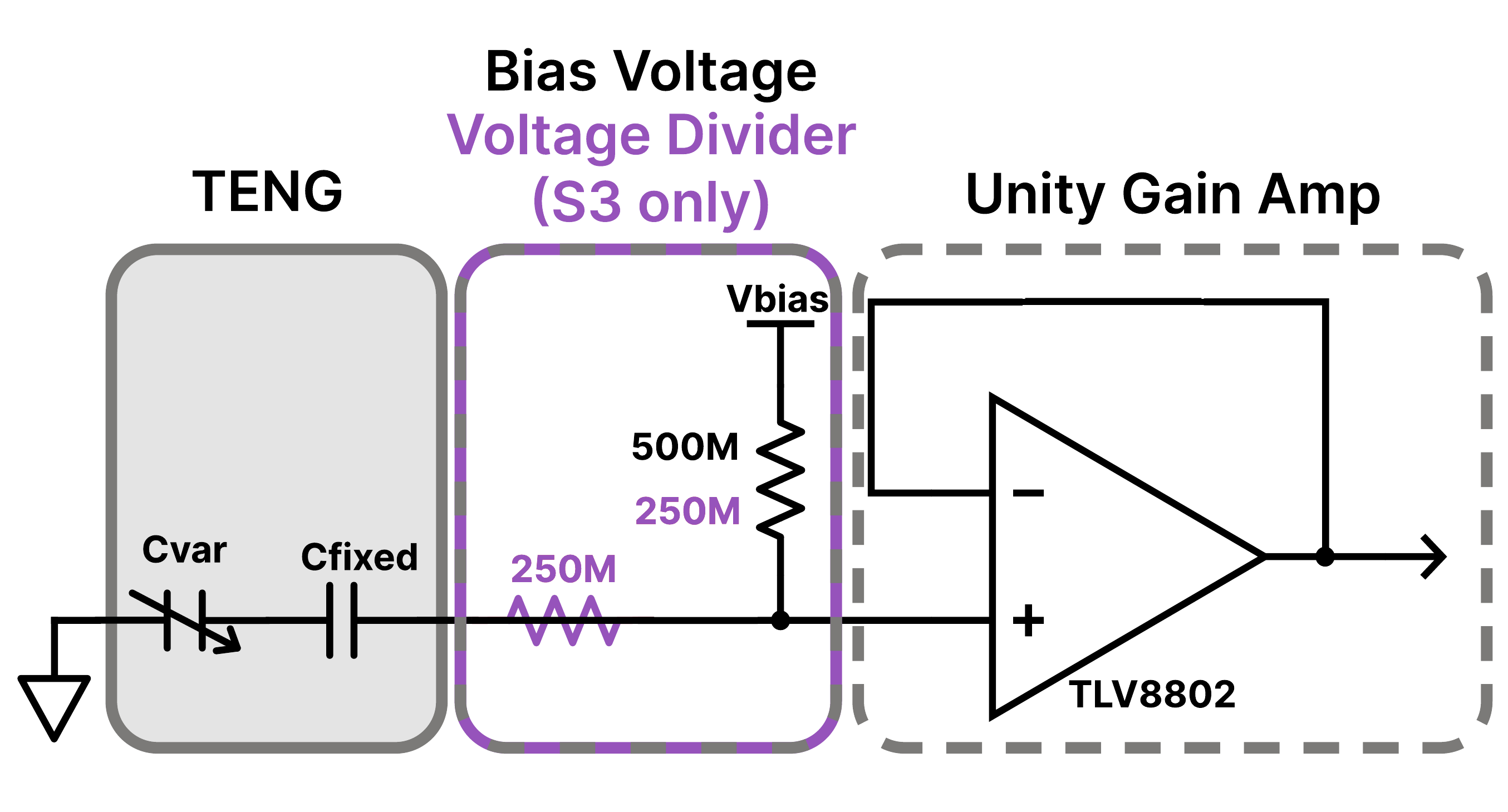}
  \vspace{-15pt}
  \caption{Analog front-end architecture. In black is the default circuit used to reduce S1's and S2's output impedance. In purple are the modifications utilized to half S3's signals.}
  \vspace{-3mm}
  \label{fig:circuit}
\end{figure}
TENGs are characterized as both having high impedance (typically in the range of megaohms) and low output current (typically in the range of hundreds of nanoamperes to microamperes), making signals from TENGs are unreadable by traditional methods ~\cite{key-parameters-teng, measurment-framework-teng}. Thus, effectively obtaining measurements from TENG sensors in glasses requires an analog front-end that lowers the outputs' impedance and uses minimal power. 


\noindent To obtain voltage measurements from our TENGs, we implement two versions of a unity gain amplifier (Figure~\ref{fig:circuit}), a circuit designed to reduce the output impedance of a signal. The unity gain amplifier utilizes the TLV8802 as its operational amplifier due to its high input impedance (18T$\Omega$) and low quiescent current ~\cite{tlv8802-datasheet}, allowing it to read from the TENG's high-impedance sources with ultra-low power consumption. The amplifier receives the TENG's output after the output adjusted by a bias voltage. This is accomplished by utilizing the TENG's internal fixed capacitance to operate as part of a high-pass filter ~\cite{teng-circuit-simulation}. S3 uses a modified circuit since its jaw signals threatened to exceed the system's voltage range of 3.3V due to the large range of forces exhibited by different users (Figure \ref{fig:fabrication_working_characterization}C). Thus, its circuit utilizes a high-impedance voltage divider to halve S3's incoming signal ~\cite{regulating_high_volt_impede_teng}. The outputs of these circuits are read by an ESP32C6 module, and the signals are logged to an on-board microSD card at a sampling rate of $100$ Hz for offline analysis; the ESP32C6 serves as a validation platform for the analog front-end and was used to collect data during our user studies.

\section{Experimental Methodology} \label{sec: expt_methodology}
 
\begin{figure}[t]
  \centering
  \includegraphics[width=1.0\linewidth]{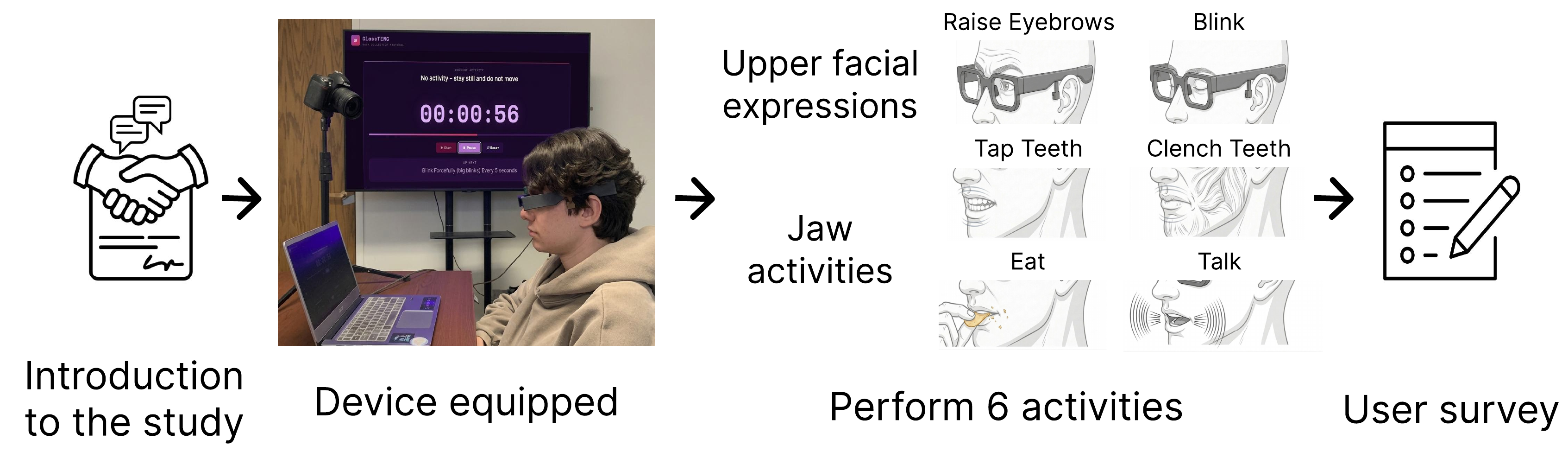}
  \vspace{-20pt}
  \caption{User study protocol of Study 1, consisting of 10 minutes of introduction and informed consent, 30 minutes of experimental procedure, and 10 minutes of survey.}
  \label{fig:user_study}
  \vspace{-5mm}
\end{figure}
\subsection{User Study Design}\label{sec:user_study_design}
We conducted two studies, one for activity recognition through the GlassTENG, and the other one to validate the heart rate (HR) from the glasses with the help of ground truth. All the study procedures were approved by the authors' Institutional Review Board (IRB), and all participants provided informed consent prior to data collection.

\noindent \textbf{Study 1 Activity Recognition:} N=20 participants ($11$ female, $9$ male) from diverse racial and ethnic backgrounds were recruited for this study to evaluate the device’s versatility across varying facial configurations. To account for prior habituation to eyewear, participants include an equal distribution (N=10) of daily glasses wearers and non-wearers. Each participant engaged in 6 different activities (not including the No Activity phase). The session started with a $120$-second baseline of No Activity, followed by a series of tasks ranging from $60$ to $180$ seconds in duration. Each activity was separated by a $30$-second rest interval. The selected activities were categorized into two primary signal suites: (1) upper facial activities, including Blinking Forcefully and Eyebrow Raises; (2) masticatory and oral movements, comprising Teeth Clenching, Teeth Tapping, Eating and Talking (Figure~\ref{fig:user_study}).
\noindent \textbf {Study 2 Pulse Validation:} To validate the pulse detected from GlassTENG, we conducted a study on heart rate monitoring  with a subset of $10$ participants. Each user remained in a seated, resting position for a duration of $5$ minutes. We used a Polar H10 chest strap \cite{polarh10} as ground-truth HR reference, which provides ECG-quality heart rate bpm accuracy \cite{polar_h10_validity}. This reference data allowed for a direct comparison against the device signals to validate GlassTENG's precision in detecting pulse signals during resting periods.

\vspace{-3mm}
\subsection{Data Processing and Evaluation} \label{sec:data_processing_eval}
Raw data from the microSD card were first digitally filtered. Based on power spectrum density analysis, noise spikes were removed with two 4th-order Butterworth bandstop filters ($20$ Hz, $40$ Hz). In addition, a $0.5$ Hz 4th-order high pass filter and a $10$ Hz 6th-order low pass filter was implemented to remove additional noise.

\noindent For Study 1, signal analysis was anchored to an initial No activity baseline for each participant. For discrete, transient activities (Blink, Raise eyebrows, Clench teeth and Tap teeth), individual events were isolated using an energy envelope detector based on the Pan-Tompkins algorithm \cite{pan-tompkins}. The remaining continuous activities applied a 2-second sliding window with $50\%$ overlap across the full signal. A fixed-length feature vector was extracted per window, comprising time and frequency-domain descriptors. To ensure signal purity, feature extraction for event-triggered classes was restricted to the isolated event windows to prevent contamination from surrounding resting-state signal.

\noindent For Study 2, we selected the optimal sensor channel for each user based on peak prominence to maximize heart rate estimation accuracy. We then use the selected sensor signal to do BPM derivation. Heart rate calculation was derived from the selected channel using a $30$-second sliding window. These values were compared with the data collected by Polar chest strap to evaluate the reliability.


\section{Results} \label{sec:results}


\subsection{Pulse Sensing from Three Sites} \label{sec: pulse_3_sites (results)}
\begin{figure}[t]
  \centering
  \includegraphics[width=1.0\linewidth]{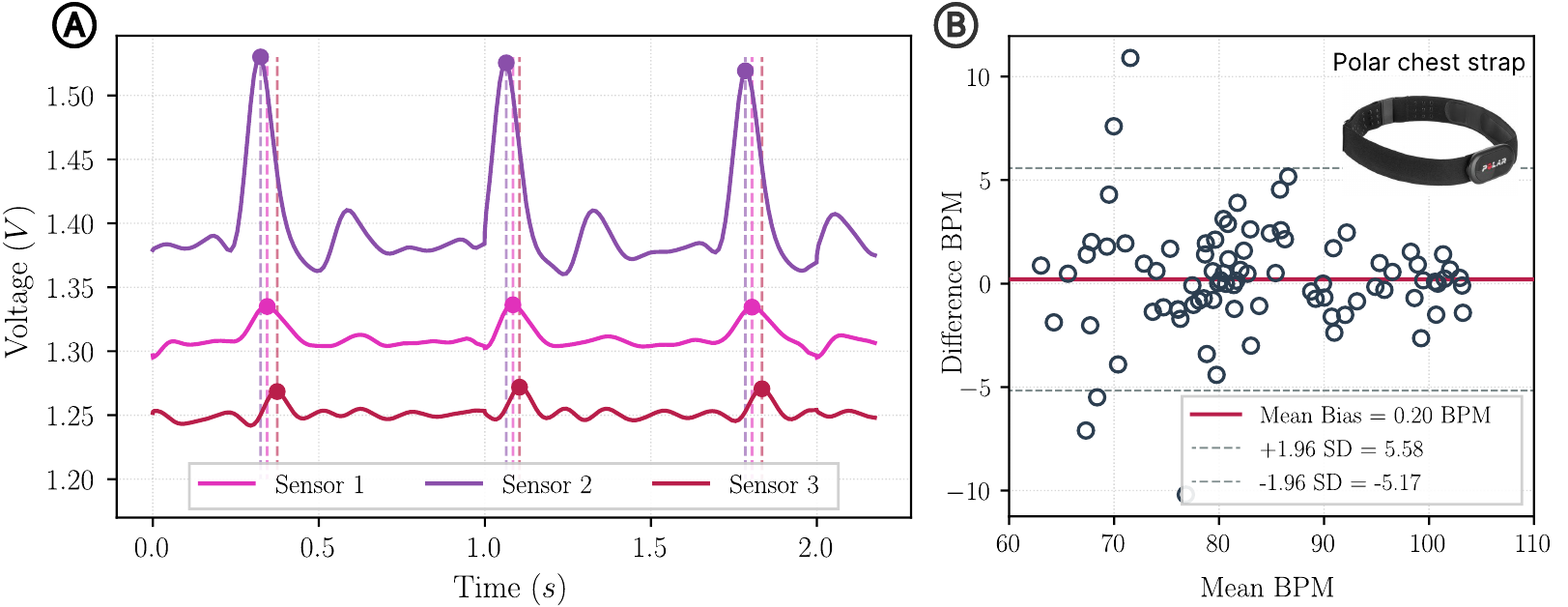}
  \vspace{-20pt}
  \caption{(A) Pulse waveforms captured simultaneously at all three GlassTENG sensor sites during No Activity, with dashed lines marking systolic peaks across sensors. (B) Bland-Altman plot comparing heart rates (BPM) from GlassTENG and the Polar chest strap ($n=10$) calculated over $30$-second windows. Top-right shows the Polar chest strap used to collect ground-truth heart rate.}
  \label{fig:pulse_bland}
  \vspace{-3mm}
\end{figure}

GlassTENG consistently captured arterial pulse waveforms at all three sensing locations (S1, S2, S3) across the 20 participants under resting conditions (Figure~\ref{fig:pulse_bland}A). Quantitative validation against the Polar chest strap ($n=10$) demonstrated a high degree of concordance between GlassTENG output and the ground truth. The system achieved a Mean Absolute Error (MAE) of $1.82$ BPM. Furthermore, the Bland-Altman analysis (Figure~\ref{fig:pulse_bland}B) revealed a mean bias of $0.20$ BPM, with the $95\%$ limits of agreement (LoA) ranging from $-5.17$ to $5.58$ BPM. These were calculated using a 30-second sliding window to maintain stable pulse estimation. The results confirm that GlassTENG's sensing capabilities are highly reliable for stationary pulse tracking.


 

\begin{figure}[t]
  \centering
  \includegraphics[width=1.0 \linewidth]{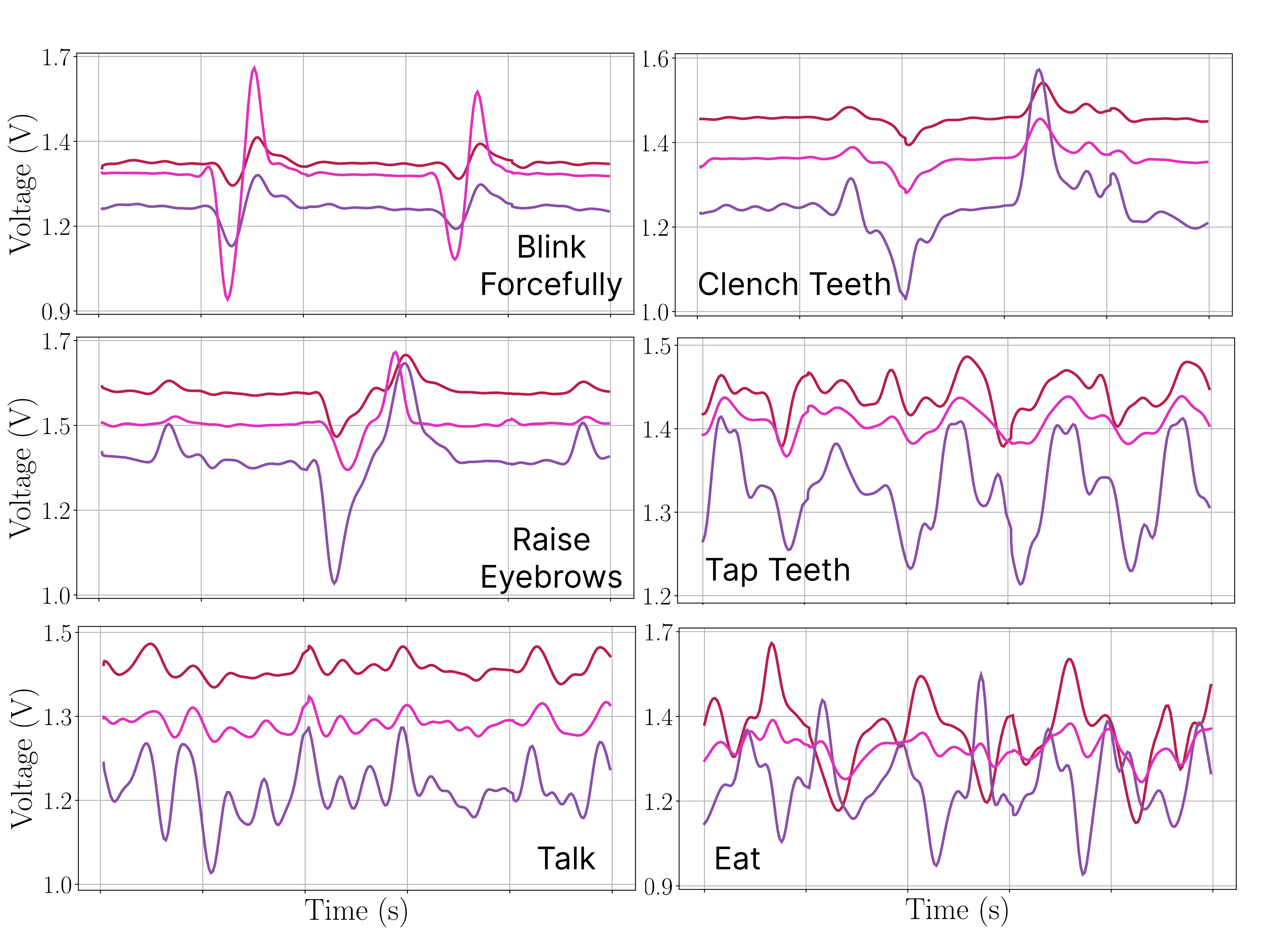}
  \vspace{-20pt}
  \caption{Signals from 3 sensors during 6 different activities in Study 1, with 2.5 seconds window size.}
  \label{fig:activities}
  \vspace{-5mm}
\end{figure}
 
\subsection{Activity Classification}\label{sec: activity classification}
\begin{figure}[t]
  \centering
  \includegraphics[width=0.95\linewidth]{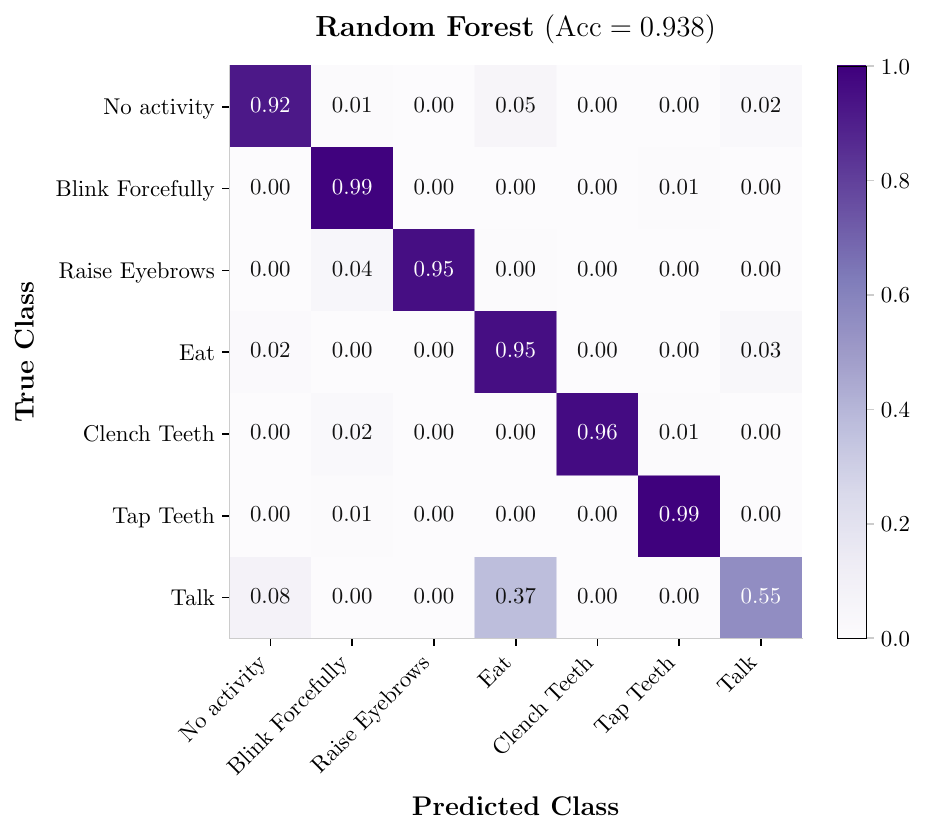}
  \vspace{-10pt}
  \caption{Normalized confusion matrix for the Random Forest classifier (LOSO, 7 classes).}
  \label{fig:confusion}
  \vspace{-5mm}
\end{figure}

Figure \ref{fig:activities} shows the signals after applying bandpass filters for each activity class from Study 1, and how each sensor has distinct peak patterns for different activities. For instance, S1 gets the highest amplitude while blinking, whereas S2 gets the higher amplitudes for the jaw activities. To evaluate classification performance, we used Leave-One-Subject-Out (LOSO) cross validation. Among the models tested (Table~\ref{tab:classification}), the Random Forest classifier achieved the highest performance with an overall accuracy of $93.8\%$. Feature importance analysis identified waveform length, spectral entropy, peak frequency, and root mean square amplitude as the most significant contributors to the model performance. The confusion matrix (Figure~\ref{fig:confusion}) shows the classification results, with Blink and Tap Teeth having the highest accuracy ($99\%$). In contrast, Talk shows the lowest accuracy ($55.0\%$), and is frequently misclassified as Eat ($37.0\%$). This is caused by the biomechanical similarities in jaw movements and temporalis muscle activation between speech and mastication. We discuss this further in Section \ref{sec: discussion}.

    
\begin{table}[t]
  \centering
  \caption{Leave-One-Subject-Out classification results (7 classes, $N=20$).}
  \label{tab:classification}
  \begin{tabular}{lcccc}
    \toprule
    \textbf{Model} & \textbf{Accuracy} & \textbf{Precision} & \textbf{Recall} & \textbf{F1} \\
    \midrule
    Random Forest  & \textbf{0.938} & \textbf{0.92} & \textbf{0.90} & \textbf{0.91}\\
    kNN            & 0.79 & 0.71 & 0.69 & 0.70 \\
    SVM (RBF)      & 0.88 & 0.85 & 0.86 & 0.85 \\
    LDA            & 0.84 & 0.79 & 0.79 & 0.78 \\
    Logistic Regression & 0.88 & 0.85 & 0.88 & 0.86 \\
    \bottomrule
    \vspace{-5mm}
  \end{tabular}
\end{table}

 
\subsection{User Experience} \label{user_exp}
Following Study 1, participants completed a structured survey covering physical comfort and social acceptability of GlassTENG's protoboard prototype. When asked about a lighter, compact version, participants reported a public comfort rating of 5.5/7, with interest highest for General Wellness Tracking (65\%) and Fitness Tracking (60\%). Adoption intent was stratified by prior eyewear experience: habitual wearers rated daily use likelihood at
5.4/7 versus 3.5/7 for non-wearers, and
4 of 5 weight-related complaints came from non-wearers — indicating that eyewear habituation significantly shapes user perception. A follow-up study with the PCB model is needed to assess acceptability at a more representative form factor.

 
\section{Discussion} \label{sec: discussion}


\subsection{Power Consumption}

GlassTENG's analog front-end surpasses similar sensor technologies in i) setting the stage for a battery-free glass platform for longitudinally deployed systems and ii) extending the overall battery life of existing glass platforms. For instance, if such a system used an 100mAh 3.3V battery to power a generic low-power MCU consuming 14mA, adding an optical PPG to the system would reduce the battery life by ~1.6-9.8 hours, a surface EMG by ~9.8-21.7 hours, and a contact microphone by ~12.2 hours (Table ~\ref{tab:sensor_comparison}). Meanwhile, GlassTENG's full 3 channel analog front-end would cost the battery \textit{less than half a minute} of its lifetime. This improvement in power consumption can contribute to bio-sensing smart glasses optimized for sustained wearability. A battery-free system may even be achieved through energy harvesting technology and by using the TENGs to power their own front-end circuitry.

\vspace{-2mm}
\subsection{Pulse and Activity Classification}
     As discussed in Section \ref{sec:results}, the readings obtained from GlassTENG's sensors and front-end circuitry was enough to both measure heart rate and perform activity recognition. Additionally, as previous work has shown with PPGs ~\cite{glabella}, GlassTENG's ability to measure the pulse from multiple locations may provide an additional pathway to obtain continuous cuffless blood pressure measurements. Combining this data with effective activity recognition could create a wearable platform with the ability to detect a number of health conditions, including stress eating, bruxism, emotional health, and potentially high blood pressure conditions such as hypertension. 

\vspace{-2mm}
\subsection{Limitations and Future Works}

The primary classification limitation is the conflation of Talking and Eating, which stems from their overlapping jaw kinematics 
at the chosen sensor sites rather than a modeling shortfall alone; richer 
temporal features and a large scale free-living study with concurrent activity combinations 
would help characterize this boundary. On the hardware side, S2's dedicated 
downward arm is the least conventional element of the frame — since S3 already 
captures overlapping vascular and jaw signals from the temple, eliminating S2 
is a natural next step toward a fully standard glasses profile. Finally, the 
current work validates the analog front-end in isolation; integrating it with 
an ultra-low-power microcontroller or custom ASIC and validating end-to-end 
energy harvesting are the remaining steps toward a fully battery-free, 
longitudinally deployed platform. This would unlock applications in dietary 
monitoring, bruxism detection, cardiovascular health, and passive 
human-computer interaction.

\vspace{-2mm}
\section{Conclusion} \label{sec:conclusion}
 
We presented GlassTENG, a triboelectric nanogenerator based sensor designed for ultra-low-power sensing of pulse, jaw, and upper facial activities on glasses. GlassTENG utilizes 3 TENG sensors with a 4.1$\mu$W combined high impedance analog front-end to collect data from across a user's face. In order to test our pulse and activity classification system, we collected data from a 20-participant user study. The users' pulse was accurately detected by all 3 sensors, and the Random Forest model used in our activity classification system achieved an overall accuracy of $93.8\%$. Additionally, users indicated that the design was publicly comfortable, and the majority were likely to use it for sustained wearability. GlassTENG's ultra-low-power, broad front-end sensing capabilities, and comfortable skin-contact paves the way for a longitudinally deployed battery-free bio-sensing platform for glasses.

\bibliographystyle{ACM-Reference-Format}
\bibliography{sample-base}

@String{Computing = "Computing" }

@String{Springer = "Springer-Verlag" }

@article{glabella,
author = {Holz, Christian and Wang, Edward J.},
title = {Glabella: Continuously Sensing Blood Pressure Behavior using an Unobtrusive Wearable Device},
year = {2017},
issue_date = {September 2017},
publisher = {Association for Computing Machinery},
address = {New York, NY, USA},
volume = {1},
number = {3},
url = {https://doi.org/10.1145/3132024},
doi = {10.1145/3132024},

month = sep,
articleno = {58},
numpages = {23}
}

@INPROCEEDINGS{pulseglasses,
  author={Constant, Nicholas and Douglas-Prawl, Orrett and Johnson, Samuel and Mankodiya, Kunal},
  booktitle={2015 IEEE 12th International Conference on Wearable and Implantable Body Sensor Networks (BSN)}, 
  title={Pulse-Glasses: An unobtrusive, wearable HR monitor with Internet-of-Things functionality}, 
  year={2015},
  volume={},
  number={},
  pages={1-5},
  doi={10.1109/BSN.2015.7299350}}

@ARTICLE{emotionglasses,
  author={Kwon, Jangho and Ha, Jihyeon and Kim, Da-Hye and Choi, Jun Won and Kim, Laehyun},
  journal={IEEE Access}, 
  title={Emotion Recognition Using a Glasses-Type Wearable Device via Multi-Channel Facial Responses}, 
  year={2021},
  volume={9},
  number={},
  pages={146392-146403},
  doi={10.1109/ACCESS.2021.3121543}}

@Article{opticalchewing,
author="Stankoski, Simon
and Kiprijanovska, Ivana
and Gjoreski, Martin
and Panchevski, Filip
and Sazdov, Borjan
and Sofronievski, Bojan
and Cleal, Andrew
and Fatoorechi, Mohsen
and Nduka, Charles
and Gjoreski, Hristijan",
title="Controlled and Real-Life Investigation of Optical Tracking Sensors in Smart Glasses for Monitoring Eating Behavior Using Deep Learning: Cross-Sectional Study",
journal="JMIR Mhealth Uhealth",
year="2024",
month="Sep",
day="26",
volume="12",
pages="e59469",
issn="2291-5222",
doi="10.2196/59469",
url="https://mhealth.jmir.org/2024/1/e59469",
url="https://doi.org/10.2196/59469",
url="http://www.ncbi.nlm.nih.gov/pubmed/39325528"
}

@inproceedings{chewingrazor,
author = {Papapanagiotou, Vasileios and Liapi, Anastasia and Delopoulos, Anastasios},
title = {Chewing Detection from Commercial Smart-glasses},
year = {2022},
isbn = {9781450395021},
publisher = {Association for Computing Machinery},
address = {New York, NY, USA},
url = {https://doi.org/10.1145/3552484.3555746},
doi = {10.1145/3552484.3555746},
booktitle = {Proceedings of the 7th International Workshop on Multimedia Assisted Dietary Management},
pages = {11–16},
numpages = {6},
location = {Lisboa, Portugal},
series = {MADiMa '22}
}

@ARTICLE{freelivingchewing,
  author={Zhang, Rui and Amft, Oliver},
  journal={IEEE Journal of Biomedical and Health Informatics}, 
  title={Monitoring Chewing and Eating in Free-Living Using Smart Eyeglasses}, 
  year={2018},
  volume={22},
  number={1},
  pages={23-32},
  doi={10.1109/JBHI.2017.2698523}}

@INPROCEEDINGS{dieteyeglasses,
  author={Zhang, Rui and Bernhart, Severin and Amft, Oliver},
  booktitle={2016 IEEE 13th International Conference on Wearable and Implantable Body Sensor Networks (BSN)}, 
  title={Diet eyeglasses: Recognising food chewing using EMG and smart eyeglasses}, 
  year={2016},
  volume={},
  number={},
  pages={7-12},
  doi={10.1109/BSN.2016.7516224}}

@inproceedings{biteglasses,
author = {Zhang, Rui and Amft, Oliver},
title = {Bite glasses: measuring chewing using emg and bone vibration in smart eyeglasses},
year = {2016},
isbn = {9781450344609},
publisher = {Association for Computing Machinery},
address = {New York, NY, USA},
url = {https://doi.org/10.1145/2971763.2971799},
doi = {10.1145/2971763.2971799},

booktitle = {Proceedings of the 2016 ACM International Symposium on Wearable Computers},
pages = {50–52},
numpages = {3},
location = {Heidelberg, Germany},
series = {ISWC '16}
}

@article{glassense,
  author    = {Chung, Jungman and Chung, Jungmin and Oh, Wonjun and Yoo, Yongkyu and Lee, Won Gu and Bang, Hyunwoo},
  title     = {A glasses-type wearable device for monitoring the patterns of food intake and facial activity},
  journal   = {Scientific Reports},
  year      = {2017},
  volume    = {7},
  number    = {1},
  pages     = {41690},
  month     = {Jan},
  doi       = {10.1038/srep41690},
  url       = {https://doi.org/10.1038/srep41690},
  issn      = {2045-2322}
}

@article{foodintakesmartglasses,
  title = {Design and Evaluation of Smart Glasses for Food Intake and Physical Activity Classification},
  author = {Chung, Jungman and
          Oh, Wonjoon and
          Baek, Dongyoub and
          Ryu, Sunwoong and
          Lee, Won Gu and
          Bang, Hyunwoo},
  year = {2018},

  number = {132},
  pages = {e56633},
  journal = {JoVE},
  doi = {doi:10.3791/56633},
}

@InProceedings{meciface,
author="Bello, Hymalai
and Suh, Sungho
and Zhou, Bo
and Lukowicz, Paul",
editor="Bravo, Jos{\'e}
and Nugent, Chris
and Cleland, Ian",
title="MeciFace: Mechanomyography and Inertial Fusion-Based Glasses for Edge Real-Time Recognition of Facial and Eating Activities",
booktitle="Proceedings of the International Conference on Ubiquitous Computing and Ambient Intelligence (UCAmI 2024)",
year="2024",
publisher="Springer Nature Switzerland",
address="Cham",
pages="393--405",

isbn="978-3-031-77571-0"
}

@inproceedings{photoreflective,
author = {Masai, Katsutoshi and Sugiura, Yuta and Ogata, Masa and Kunze, Kai and Inami, Masahiko and Sugimoto, Maki},
title = {Facial Expression Recognition in Daily Life by Embedded Photo Reflective Sensors on Smart Eyewear},
year = {2016},
isbn = {9781450341370},
publisher = {Association for Computing Machinery},
address = {New York, NY, USA},
url = {https://doi.org/10.1145/2856767.2856770},
doi = {10.1145/2856767.2856770},

pages = {317–326},
numpages = {10},
location = {Sonoma, California, USA},
series = {IUI '16}
}

@article{acousticfacial,
author = {Xie, Wentao and Zhang, Qian and Zhang, Jin},
title = {Acoustic-based Upper Facial Action Recognition for Smart Eyewear},
year = {2021},
issue_date = {June 2021},
publisher = {Association for Computing Machinery},
address = {New York, NY, USA},
volume = {5},
number = {2},
url = {https://doi.org/10.1145/3448105},
doi = {10.1145/3448105},

journal = {Proc. ACM Interact. Mob. Wearable Ubiquitous Technol.},
month = jun,
articleno = {41},
numpages = {28}
}

@article{selfpoweredeye,
title = {Self-powered eye-computer interaction via a triboelectric nanogenerator},
journal = {Device},
volume = {2},
number = {1},
pages = {100252},
year = {2024},
issn = {2666-9986},
doi = {https://doi.org/10.1016/j.device.2023.100252},
url = {https://www.sciencedirect.com/science/article/pii/S266699862300409X},
author = {Junyi Yin and Vishesh Kashyap and Shaolei Wang and Xiao Xiao and Trinny Tat and Jun Chen}
}

@article{actsonic,
author = {Mahmud, Saif and Parikh, Vineet and Liang, Qikang and Li, Ke and Zhang, Ruidong and Ajit, Ashwin and Gunda, Vipin and Agarwal, Devansh and Guimbretiere, Francois and Zhang, Cheng},
title = {ActSonic: Recognizing Everyday Activities from Inaudible Acoustic Wave Around the Body},
year = {2024},
issue_date = {December 2024},
publisher = {Association for Computing Machinery},
address = {New York, NY, USA},
volume = {8},
number = {4},
url = {https://doi.org/10.1145/3699752},
doi = {10.1145/3699752},
journal = {Proc. ACM Interact. Mob. Wearable Ubiquitous Technol.},
month = nov,
articleno = {183},
numpages = {32}
}

@article{qu2025,
  author    = {Qu, Xuecheng and Wan, Jiahao and Zhao, Haohan and Xu, Shunyuan and Cheng, Xiangrong and Yang, Bo and Li, Zhibin and Ji, Linhong and Wu, Jinting and Li, Zhou and Cheng, Jia and Li, Chong},
  title     = {Closed-loop wearable neurostimulation system with triboelectric sensing to alleviate hemifacial spasms},
  journal   = {Nature Communications},
  year      = {2026},
  volume    = {16},
  number    = {1},
  pages     = {11148},
  month     = {Jan},
  doi       = {10.1038/s41467-025-67121-9},
  url       = {https://doi.org/10.1038/s41467-025-67121-9},
  issn      = {2041-1723}
}

@inproceedings{clip_free_glabella_predecesor,
  author    = {Zheng, Yali and Leung, Billy and Sy, Stanley and Zhang, Yuanting and Poon, Carmen C. Y.},
  title     = {A clip-free eyeglasses-based wearable monitoring device for measuring photoplethysmograhic signals},
  booktitle = {2012 Annual International Conference of the IEEE Engineering in Medicine and Biology Society},
  year      = {2012},
  pages     = {5022--5025},
  doi       = {10.1109/EMBC.2012.6347121},
  issn      = {2694-0604}
}

@Article{BP_based_on_PTT_PAT_review,
AUTHOR = {Zhou, Zi-Bo and Cui, Tian-Rui and Li, Ding and Jian, Jin-Ming and Li, Zhen and Ji, Shou-Rui and Li, Xin and Xu, Jian-Dong and Liu, Hou-Fang and Yang, Yi and Ren, Tian-Ling},
TITLE = {Wearable Continuous Blood Pressure Monitoring Devices Based on Pulse Wave Transit Time and Pulse Arrival Time: A Review},
JOURNAL = {Materials},
VOLUME = {16},
YEAR = {2023},
NUMBER = {6},
ARTICLE-NUMBER = {2133},
URL = {https://www.mdpi.com/1996-1944/16/6/2133},
PubMedID = {36984013},
ISSN = {1996-1944},
DOI = {10.3390/ma16062133}
}

@article{application_hypertension_using_ppg,
  author    = {Elgendi, Mohamed and Fletcher, Richard and Liang, Yongbo stone and Howard, Newton and Lovell, Nigel H. and Abbott, Derek and Lim, Kenneth and Ward, Rabab},
  title     = {The use of photoplethysmography for assessing hypertension},
  journal   = {npj Digital Medicine},
  year      = {2019},
  volume    = {2},
  number    = {1},
  pages     = {60},
  month     = {Jun},
  doi       = {10.1038/s41746-019-0136-7},
  url       = {https://doi.org/10.1038/s41746-019-0136-7},
  issn      = {2398-6352}
}

@inproceedings{non_table_capglasses,
author = {Matthies, Denys J.C. and Weerasinghe, Chamod and Urban, Bodo and Nanayakkara, Suranga},
title = {CapGlasses: Untethered Capacitive Sensing with Smart Glasses},
year = {2021},
isbn = {9781450384285},
publisher = {Association for Computing Machinery},
address = {New York, NY, USA},
url = {https://doi.org/10.1145/3458709.3458945},
doi = {10.1145/3458709.3458945},
booktitle = {Proceedings of the Augmented Humans International Conference 2021},
pages = {121–130},
numpages = {10},
location = {Rovaniemi, Finland},
series = {AHs '21}
}

@inproceedings{non_table_tunnelvision,
author = {Zhang, Qing and Yamamura, Hiroo and Baldauf, Holger and Zheng, Dingding and Chen, Kanyu and Yamaoka, Junichi and Kunze, Kai},
title = {Tunnel Vision – Dynamic Peripheral Vision Blocking Glasses for Reducing Motion Sickness Symptoms},
year = {2021},
isbn = {9781450384629},
publisher = {Association for Computing Machinery},
address = {New York, NY, USA},
url = {https://doi.org/10.1145/3460421.3478824},
doi = {10.1145/3460421.3478824},
booktitle = {Proceedings of the 2021 ACM International Symposium on Wearable Computers},
pages = {48–52},
numpages = {5},
location = {Virtual, USA},
series = {ISWC '21}
}

@misc{visioncouncil2021,
  author       = {{The Vision Council}},
  title        = {Vision Council of {America}: {VisionWatch} Consumer Research},
  year         = {2021},
  howpublished = {\url{https://thevisioncouncil.org}},
  note         = {Accessed May 2026}
}

@techreport{who2019vision,
  author      = {{World Health Organization}},
  title       = {World Report on Vision},
  institution = {World Health Organization},
  year        = {2019},
  url         = {https://www.who.int/publications/i/item/9789241516570}
}

@misc{Subin2026,
  author       = {Subin, Samantha},
  title        = {Ray-Ban maker EssilorLuxottica says it more than tripled Meta AI glasses sales in 2025},
  howpublished = {CNBC},
  year         = {2026},
  month        = {Feb},
  day          = {11},
  url          = {https://www.cnbc.com/2026/02/11/ray-ban-maker-essilorluxottica-triples-sales-of-meta-ai-glasses.html}
}

@article{wearable_abandonment,
title = {Abandonment of personal quantification: A review and empirical study investigating reasons for wearable activity tracking attrition},
journal = {Computers in Human Behavior},
volume = {102},
pages = {223-237},
year = {2020},
issn = {0747-5632},
doi = {https://doi.org/10.1016/j.chb.2019.08.025},
url = {https://www.sciencedirect.com/science/article/pii/S0747563219303127},
author = {Christiane Attig and Thomas Franke},

}

@article{teng_original,
title = {Flexible triboelectric generator},
journal = {Nano Energy},
volume = {1},
number = {2},
pages = {328-334},
year = {2012},
issn = {2211-2855},
doi = {https://doi.org/10.1016/j.nanoen.2012.01.004},
url = {https://www.sciencedirect.com/science/article/pii/S2211285512000481},
author = {Feng-Ru Fan and Zhong-Qun Tian and Zhong {Lin Wang}},
}

@article{Teng_new_energy,
author = {Wu, Changsheng and Wang, Aurelia C. and Ding, Wenbo and Guo, Hengyu and Wang, Zhong Lin},
title = {Triboelectric Nanogenerator: A Foundation of the Energy for the New Era},
journal = {Advanced Energy Materials},
volume = {9},
number = {1},
pages = {1802906},
doi = {https://doi.org/10.1002/aenm.201802906},
url = {https://advanced.onlinelibrary.wiley.com/doi/abs/10.1002/aenm.201802906},
eprint = {https://advanced.onlinelibrary.wiley.com/doi/pdf/10.1002/aenm.201802906},
year = {2019}
}

@article{teng_sp_new_energy,
  author    = {Wang, Zhong Lin},
  title     = {Triboelectric Nanogenerators as New Energy Technology for Self-Powered Systems and as Active Mechanical and Chemical Sensors},
  journal   = {ACS Nano},
  volume    = {7},
  number    = {11},
  pages     = {9533--9557},
  year      = {2013},
  month     = {November},
  publisher = {American Chemical Society},
  doi       = {10.1021/nn404614z},
  url       = {https://doi.org/10.1021/nn404614z},
  issn      = {1936-0851}
}

@Article{teng_progress,
author ="Wang, Zhong Lin and Chen, Jun and Lin, Long",
title  ="Progress in triboelectric nanogenerators as a new energy technology and self-powered sensors",
journal  ="Energy Environ. Sci.",
year  ="2015",
volume  ="8",
issue  ="8",
pages  ="2250-2282",
publisher  ="The Royal Society of Chemistry",
doi  ="10.1039/C5EE01532D",
url  ="http://dx.doi.org/10.1039/C5EE01532D",
}

@ARTICLE{key-parameters-teng,
  author={Fachechi, Luca and Blasi, Laura and Mastronardi, Vincenzo Mariano and De Vittorio, Massimo and Todaro, Maria Teresa},
  journal={IEEE Transactions on Instrumentation and Measurement}, 
  title={Effective and Accurate Approach for Measuring Key Parameters in Triboelectric Nanogenerators}, 
  year={2023},
  volume={72},
  number={},
  pages={1-8},
  doi={10.1109/TIM.2023.3328701}}

@article{measurment-framework-teng,
title = {A configurable high-precision multi-parameter signal measurement method and circuit framework for triboelectric nanogenerator characterization},
journal = {Nano Energy},
volume = {141},
pages = {111107},
year = {2025},
issn = {2211-2855},
doi = {https://doi.org/10.1016/j.nanoen.2025.111107},
url = {https://www.sciencedirect.com/science/article/pii/S2211285525004665},
author = {Xingxu Jiang and Meng Chen and Wenqiu Liu and Kecen Li and Shiwei Xu and Hua Yu},

}

@article{teng-circuit-simulation,
title = {Triboelectric nanogenerator module for circuit design and simulation},
journal = {Nano Energy},
volume = {107},
pages = {108139},
year = {2023},
issn = {2211-2855},
doi = {https://doi.org/10.1016/j.nanoen.2022.108139},
url = {https://www.sciencedirect.com/science/article/pii/S2211285522012174},
author = {Kun Wang and Yitao Liao and Wenhao Li and Yongai Zhang and Xiongtu Zhou and Chaoxing Wu and Rong Chen and Tae Whan Kim},

}

@article{facial_emg_nature,
  author    = {Gjoreski, Martin and Kiprijanovska, Ivana and Stankoski, Simon and Mavridou, Ifigeneia and Broulidakis, M. John and Gjoreski, Hristijan and Nduka, Charles},
  title     = {Facial EMG sensing for monitoring affect using a wearable device},
  journal   = {Scientific Reports},
  year      = {2022},
  volume    = {12},
  number    = {1},
  pages     = {16876},
  month     = {Oct},
  doi       = {10.1038/s41598-022-21456-1},
  url       = {https://doi.org/10.1038/s41598-022-21456-1},
  issn      = {2045-2322}
}

@article{saturn_original,
author = {Arora, Nivedita and Zhang, Steven L. and Shahmiri, Fereshteh and Osorio, Diego and Wang, Yi-Cheng and Gupta, Mohit and Wang, Zhengjun and Starner, Thad and Wang, Zhong Lin and Abowd, Gregory D.},
title = {SATURN: A Thin and Flexible Self-powered Microphone Leveraging Triboelectric Nanogenerator},
year = {2018},
issue_date = {June 2018},
publisher = {Association for Computing Machinery},
address = {New York, NY, USA},
volume = {2},
number = {2},
url = {https://doi.org/10.1145/3214263},
doi = {10.1145/3214263},
journal = {Proc. ACM Interact. Mob. Wearable Ubiquitous Technol.},
month = jul,
articleno = {60},
numpages = {28}
}

@Article{facial_emg_ex2,
AUTHOR = {Gul, Jahan Zeb and Fatima, Noor and Mohy Ud Din, Zia and Khan, Maryam and Kim, Woo Young and Rehman, Muhammad Muqeet},
TITLE = {Advanced Sensing System for Sleep Bruxism across Multiple Postures via EMG and Machine Learning},
JOURNAL = {Sensors},
VOLUME = {24},
YEAR = {2024},
NUMBER = {16},
ARTICLE-NUMBER = {5426},
URL = {https://www.mdpi.com/1424-8220/24/16/5426},
PubMedID = {39205120},
ISSN = {1424-8220},
DOI = {10.3390/s24165426}
}

@article{eyegesturelistener,
author = {Sun, Tao and Zhao, Yankai and Xie, Wentao and Li, Jiao and Ma, Yongyu and Zhang, Jin},
title = {EyeGesener: Eye Gesture Listener for Smart Glasses Interaction Using Acoustic Sensing},
year = {2024},
issue_date = {September 2024},
publisher = {Association for Computing Machinery},
address = {New York, NY, USA},
volume = {8},
number = {3},
url = {https://doi.org/10.1145/3678541},
doi = {10.1145/3678541},
journal = {Proc. ACM Interact. Mob. Wearable Ubiquitous Technol.},
month = sep,
articleno = {128},
numpages = {28}
}

@article{ppg_limitations,
  author    = {Fine, Jesse and Branan, Kimberly L. and Rodriguez, Andres J. and Boonya-Ananta, Tananant and Ajmal and Ramella-Roman, Jessica C. and McShane, Michael J. and Cot{\'e}, Gerard L.},
  title     = {Sources of Inaccuracy in Photoplethysmography for Continuous Cardiovascular Monitoring},
  journal   = {Biosensors},
  year      = {2021},
  volume    = {11},
  number    = {4},
  pages     = {126},
  month     = {Apr},
  doi       = {10.3390/bios11040126},
  url       = {https://doi.org/10.3390/bios11040126},
  issn      = {2079-6374}
}

@article{Hertzman1937,
author = {Hertzman, A B},
address = {[Maywood, N.J.] :},
issn = {1535-3702},
journal = {Experimental biology and medicine.},
lccn = {2001211112},
number = {3},
publisher = {Society for Experimental Biology and Medicine,},
title = {Photoelectric Plethysmography of the Fingers and Toes in Man},
volume = {37},
year = {1937-12-01},
}

@article{eyebrow_as_communication,
  author    = {Nota, Naomi and Trujillo, James P. and Jacobs, Vere and Holler, Judith},
  title     = {Facilitating question identification through natural intensity eyebrow movements in virtual avatars},
  journal   = {Scientific Reports},
  year      = {2023},
  volume    = {13},
  number    = {1},
  pages     = {21295},
  month     = {Dec},
  doi       = {10.1038/s41598-023-48586-4},
  url       = {https://doi.org/10.1038/s41598-023-48586-4},
  issn      = {2045-2322}
}

@article{eye_and_fatigue_nature,
  author    = {Zargari Marandi, Ramtin and Madeleine, Pascal and Omland, {\O}yvind and Vuillerme, Nicolas and Samani, Afshin},
  title     = {Eye movement characteristics reflected fatigue development in both young and elderly individuals},
  journal   = {Scientific Reports},
  year      = {2018},
  volume    = {8},
  number    = {1},
  pages     = {13148},
  month     = {Sep},
  doi       = {10.1038/s41598-018-31577-1},
  url       = {https://doi.org/10.1038/s41598-018-31577-1},
  issn      = {2045-2322}
}

@article{blink_and_cognitive_load,
  author    = {Alyan, Emad and Arnau, Stefan and Reiser, Julian Elias and Getzmann, Stephan and Karthaus, Melanie and Wascher, Edmund},
  title     = {Blink-related EEG activity measures cognitive load during proactive and reactive driving},
  journal   = {Scientific Reports},
  year      = {2023},
  volume    = {13},
  number    = {1},
  pages     = {19379},
  month     = {Nov},
  doi       = {10.1038/s41598-023-46738-0},
  url       = {https://doi.org/10.1038/s41598-023-46738-0},
  issn      = {2045-2322}
}

@article{teng_wrist_pulse,
  author    = {V, Karthikeyan and S, Vivekanandan},
  title     = {IoT-based triboelectric nanogenerator for wrist pulse acquisition and analysis},
  journal   = {RSC Advances},
  year      = {2025},
  volume    = {15},
  number    = {5},
  pages     = {3592--3601},
  month     = {Jan},
  doi       = {10.1039/d4ra08200a},
  url       = {https://doi.org/10.1039/d4ra08200a},
  issn      = {2046-2069}
}

@article{teng_respiration,
  author    = {Xu, Hongqiang and Han, Weiqiao and Yuce, Mehmet Rasit},
  title     = {A Wearable Device with Triboelectric Nanogenerator Sensing for Respiration and Spirometry Monitoring},
  journal   = {ACS Sensors},
  year      = {2025},
  volume    = {10},
  number    = {1},
  pages     = {264--271},
  month     = {Jan},
  doi       = {10.1021/acssensors.4c02350},
  url       = {https://doi.org/10.1021/acssensors.4c02350},
  issn      = {2379-3694}
}

@Article{teng_muscle,
AUTHOR = {Liu, Jing and Zhang, Yi and Liu, Xia and Sun, Chenming and Wang, Youquan},
TITLE = {Muscle Strength Training and Monitoring Device Based on Triboelectric Nanogenerator for Knee Joint Surgery},
JOURNAL = {Micromachines},
VOLUME = {16},
YEAR = {2025},
NUMBER = {12},
ARTICLE-NUMBER = {1387},
URL = {https://www.mdpi.com/2072-666X/16/12/1387},
PubMedID = {41470552},
ISSN = {2072-666X},

DOI = {10.3390/mi16121387}
}

@Article{gait_teng,
author ="Parashar, Parag and Sharma, Manish Kumar and Nahak, Bishal Kumar and Khan, Arshad and Hsu, Wei-Zan and Tseng, Yao-Hsuan and Chowdhury, Jaba Roy and Huang, Yu-Hui and Liao, Jen-Chung and Kao, Fu-Cheng and Lin, Zong-Hong",
title  ="Machine learning-driven gait-assisted self-powered wearable sensing: a triboelectric nanogenerator-based advanced healthcare monitoring",
journal  ="J. Mater. Chem. A",
year  ="2025",
volume  ="13",
issue  ="19",
pages  ="13750-13762",
publisher  ="The Royal Society of Chemistry",
doi  ="10.1039/D4TA07496C",
url  ="http://dx.doi.org/10.1039/D4TA07496C",
}

@article{teng_nanopore_pulse,
author = {Zhang, Tao and Yao, Chuanjie and Xu, Xingyuan and Liu, Zhibo and Liu, Zhengjie and Sun, Tiancheng and Huang, Shuang and Huang, Xinshuo and Farah, Shady and Shi, Peng and Chen, Hui-jiuan and Xie, Xi},
title = {Nanopores-templated CNT/PDMS Microcolumn Substrate for the Fabrication of Wearable Triboelectric Nanogenerator Sensors to Monitor Human Pulse and Blood Pressure},
journal = {Advanced Materials Technologies},
volume = {10},
number = {2},
pages = {2400749},
doi = {https://doi.org/10.1002/admt.202400749},
url = {https://advanced.onlinelibrary.wiley.com/doi/abs/10.1002/admt.202400749},
eprint = {https://advanced.onlinelibrary.wiley.com/doi/pdf/10.1002/admt.202400749},

year = {2025}
}

@article{internet_of_batteryless_things,
author = {Ahmed, Saad and Islam, Bashima and Yildirim, Kasim Sinan and Zimmerling, Marco and Pawe\l{}czak, Przemys\l{}aw and Alizai, Muhammad Hamad and Lucia, Brandon and Mottola, Luca and Sorber, Jacob and Hester, Josiah},
title = {The Internet of Batteryless Things},
year = {2024},
issue_date = {March 2024},
publisher = {Association for Computing Machinery},
address = {New York, NY, USA},
volume = {67},
number = {3},
issn = {0001-0782},
url = {https://doi.org/10.1145/3624718},
doi = {10.1145/3624718},
journal = {Commun. ACM},
month = feb,
pages = {64–73},
numpages = {10}
}

@article{teng_for_machine_monitoring,
author = {Zhao, Hongfa and Shu, Mingrui and Ai, Zihao and Lou, Zirui and Sou, Kit Wa and Lu, Chengyue and Jin, Yuchao and Wang, Zihan and Wang, Jiyu and Wu, Changsheng and Cao, Yidan and Xu, Xiaomin and Ding, Wenbo},
title = {A Highly Sensitive Triboelectric Vibration Sensor for Machinery Condition Monitoring},
journal = {Advanced Energy Materials},
volume = {12},
number = {37},
pages = {2201132},
doi = {https://doi.org/10.1002/aenm.202201132},
url = {https://advanced.onlinelibrary.wiley.com/doi/abs/10.1002/aenm.202201132},
eprint = {https://advanced.onlinelibrary.wiley.com/doi/pdf/10.1002/aenm.202201132},

year = {2022}
}

@inproceedings{battery_free_eye_glasses,
author = {Li, Tianxing and Zhou, Xia},
title = {Battery-Free Eye Tracker on Glasses},
year = {2018},
isbn = {9781450359030},
publisher = {Association for Computing Machinery},
address = {New York, NY, USA},
url = {https://doi.org/10.1145/3241539.3241578},
doi = {10.1145/3241539.3241578},
booktitle = {Proceedings of the 24th Annual International Conference on Mobile Computing and Networking},
pages = {67–82},
numpages = {16},
location = {New Delhi, India},
series = {MobiCom '18}
}

@Article{self_powered_PDms,
AUTHOR = {Wang, Jie and Qian, Shuo and Yu, Junbin and Zhang, Qiang and Yuan, Zhongyun and Sang, Shengbo and Zhou, Xiaohong and Sun, Lining},
TITLE = {Flexible and Wearable PDMS-Based Triboelectric Nanogenerator for Self-Powered Tactile Sensing},
JOURNAL = {Nanomaterials},
VOLUME = {9},
YEAR = {2019},
NUMBER = {9},
ARTICLE-NUMBER = {1304},
URL = {https://www.mdpi.com/2079-4991/9/9/1304},
PubMedID = {31547316},
ISSN = {2079-4991},
DOI = {10.3390/nano9091304}
}

@article{self_power_piezo,
author = {Wang, Wei and Cheng, Bingnan and Feng, Wuwei and He, Bin and Liu, Shuo},
title = {High-performance flexible piezoelectric nanogenerator with folded structure based on CNTs-modified BC$\beta$ZT/P(VDF-HFP) composite film},
journal = {Journal of Applied Polymer Science},
volume = {140},
number = {32},
pages = {e54261},
doi = {https://doi.org/10.1002/app.54261},
url = {https://onlinelibrary.wiley.com/doi/abs/10.1002/app.54261},
eprint = {https://onlinelibrary.wiley.com/doi/pdf/10.1002/app.54261},

year = {2023}
}

@article{self_power_thermo,
  author    = {Feng, Rui and Tang, Fei and Zhang, Ning and Wang, Xiaohao},
  title     = {Flexible, High-Power Density, Wearable Thermoelectric Nanogenerator and Self-Powered Temperature Sensor},
  journal   = {ACS Applied Materials \& Interfaces},
  year      = {2019},
  volume    = {11},
  number    = {42},
  pages     = {38616--38624},
  month     = {Oct},
  doi       = {10.1021/acsami.9b11435},
  url       = {https://doi.org/10.1021/acsami.9b11435},
  publisher = {American Chemical Society},
  issn      = {1944-8244}
}

@inproceedings{texteng,
author = {Batra, Ritik and Pourjafarian, Narjes and Chang, Samantha and Tsai, Margaret and Revelo, Jacob and Kao, Cindy Hsin-Liu},
title = {texTENG: Fabricating Wearable Textile-Based Triboelectric Nanogenerators},
year = {2025},
isbn = {9798400715662},
publisher = {Association for Computing Machinery},
address = {New York, NY, USA},
url = {https://doi.org/10.1145/3745900.3746071},
doi = {10.1145/3745900.3746071},
booktitle = {Proceedings of the Augmented Humans International Conference 2025},
pages = {124–138},
numpages = {15},
location = {
},
series = {AHs '25}
}

@inproceedings{mars_original,
author = {Arora, Nivedita and Mirzazadeh, Ali and Moon, Injoo and Ramey, Charles and Zhao, Yuhui and Rodriguez, Daniela C. and Abowd, Gregory D. and Starner, Thad},
title = {MARS: Nano-Power Battery-free Wireless Interfaces for Touch, Swipe and Speech Input},
year = {2021},
isbn = {9781450386357},
publisher = {Association for Computing Machinery},
address = {New York, NY, USA},
url = {https://doi.org/10.1145/3472749.3474823},
doi = {10.1145/3472749.3474823},
booktitle = {The 34th Annual ACM Symposium on User Interface Software and Technology},
pages = {1305–1325},
numpages = {21},
location = {Virtual Event, USA},
series = {UIST '21}
}

@article{regulating_high_volt_impede_teng,
title = {Regulating the high-voltage and high-impedance characteristics of triboelectric nanogenerator toward practical self-powered sensors},
journal = {Nano Energy},
volume = {87},
pages = {106137},
year = {2021},
issn = {2211-2855},
doi = {https://doi.org/10.1016/j.nanoen.2021.106137},
url = {https://www.sciencedirect.com/science/article/pii/S2211285521003931},
author = {Shan Lu and Wenqian Lei and Lingxiao Gao and Xin Chen and Daqiao Tong and Pengfei Yuan and Xiaojing Mu and Hua Yu}
}

@ARTICLE{pan-tompkins,
  author={Pan, Jiapu and Tompkins, Willis J.},
  journal={IEEE Transactions on Biomedical Engineering}, 
  title={A Real-Time QRS Detection Algorithm}, 
  year={1985},
  volume={BME-32},
  number={3},
  pages={230-236},
  doi={10.1109/TBME.1985.325532}}

@misc{polarh10,
  author       = {{Polar Electro}},
  title        = {Polar {H10} Heart Rate Sensor},
  year         = {2023},
  howpublished = {\url{https://www.polar.com/us-en/sensors/h10-heart-rate-sensor/}},
}

@article{polar_h10_validity,
  author    = {Chung, Victor and Chopin, Louise and Karadayi, Julien and Gr{\`e}zes, Julie},
  title     = {Validity of the Polar H10 for Continuous Measures of Heart Rate and Heart Rate Synchrony Analysis},
  journal   = {Sensors},
  year      = {2026},
  volume    = {26},
  number    = {3},
  pages     = {855},
  month     = {Jan},
  doi       = {10.3390/s26030855},
  url       = {https://doi.org/10.3390/s26030855},
  issn      = {1424-8220}
}

@article{starner2002challenges,
  title={The challenges of wearable computing: Part 1},
  author={Starner, Thad},
  journal={Ieee Micro},
  volume={21},
  number={4},
  pages={44--52},
  year={2002},
  publisher={IEEE}
}

@article{indoor_solar_harvesting, author = {Saeed, Muhammad Adnan and Kim, Sang Hyeon and Baek, Kiwook and Kim, Jong-Woon and Kim, Joo Hyun and Lee, Sang Kyu and Kim, Han Young}, title = {Indoor Photovoltaic Energy Harvesting Based on Semiconducting {$\pi$}-Conjugated Polymers and Oligomeric Materials toward Future {IoT} Applications}, journal = {Polymer Journal}, volume = {54}, number = {12}, pages = {1469--1490}, year = {2022}, doi = {10.1038/s41428-022-00727-8} }

@inproceedings{energy_harvesing_grosse,
  title={Exploring the design space for energy-harvesting situated displays},
  author={Grosse-Puppendahl, Tobias and Hodges, Steve and Chen, Nicholas and Helmes, John and Taylor, Stuart and Scott, James and Fromm, Josh and Sweeney, David},
  booktitle={Proceedings of the 29th Annual Symposium on User Interface Software and Technology},
  pages={41--48},
  year={2016}
}

@manual{msp430-datasheet,
  title        = {{MSP430FR2x3x} Mixed-Signal Microcontroller Datasheet},
  author       = {{Texas Instruments}},
  year         = {2023},
  url          = {https://www.ti.com/product/MSP430FR2433}
}

@article{battery_bottle_neck,
  author    = {Wang, Lie and Zhang, Ye and Bruce, Peter G.},
  title     = {Batteries for wearables},
  journal   = {National Science Review},
  volume    = {10},
  number    = {1},
  pages     = {nwac062},
  year      = {2023},
  month     = {January},
  publisher = {Oxford University Press},
  doi       = {10.1093/nsr/nwac062},
  pmid      = {36684516},
  pmcid     = {PMC9843125}
}

@article{batt_recharge,
    author = {Yin, Lu and Wang, Joseph},
    title = {Wearable energy systems: what are the limits and limitations?},
    journal = {National Science Review},
    volume = {10},
    number = {1},
    pages = {nwac060},
    year = {2023},
    month = {01},
    issn = {2095-5138},
    doi = {10.1093/nsr/nwac060},
    url = {https://doi.org/10.1093/nsr/nwac060},
    eprint = {https://academic.oup.com/nsr/article-pdf/10/1/nwac060/48727996/nwac060.pdf},
}

@manual{tlv8802-datasheet,
  title        = {TLV8801/TLV8802 320 nA Nanopower Operational Amplifiers for Cost-Optimized Systems},
  author       = {Texas Instruments},
  year         = {2016},
  url          = {https://www.ti.com/lit/ds/symlink/tlv8802.pdf?ts=1779602202835&ref_url=https%253A%252F%252Fwww.ti.com%252Fproduct%252FTLV8802}
}

@article{vanhelleputte2015soc,
  author  = {Van Helleputte, Nick and Konijnenburg, Mario and Pettine, Jacopo
             and Jee, Dong-Woo and Kim, Hyejung and Morgado, Alonso and
             Van Wegberg, Roland and Torfs, Tom and Mohan, Refet and
             Breeschoten, Arjan and Van Hoof, Chris and Yazicioglu, Refet Firat},
  title   = {A 345 {$\mu$W} Multi-Sensor Biomedical {SoC} With Bio-Impedance,
             3-Channel {ECG}, Motion Artifact Reduction, and Integrated {DSP}},
  journal = {IEEE Journal of Solid-State Circuits},
  volume  = {50},
  number  = {1},
  pages   = {230--244},
  year    = {2015},
  doi     = {10.1109/JSSC.2014.2359962}
}

@inproceedings{battery_free_hd_streaming,
author = {Naderiparizi, Saman and Hessar, Mehrdad and Talla, Vamsi and Gollakota, Shyamnath and Smith, Joshua R.},
title = {Towards battery-free HD video streaming},
year = {2018},
isbn = {9781931971430},
publisher = {USENIX Association},
address = {USA},
booktitle = {Proceedings of the 15th USENIX Conference on Networked Systems Design and Implementation},
pages = {233–247},
numpages = {15},
location = {Renton, WA, USA},
series = {NSDI'18}
}

\end{document}